\documentclass[conference]{IEEEtran}
\IEEEoverridecommandlockouts

\usepackage{cite}
\usepackage{amsmath,amssymb,amsfonts}
\usepackage{algorithmic}
\usepackage{graphicx}
\usepackage{textcomp}
\usepackage{xcolor}
\usepackage{bm}
\usepackage{color}
\usepackage{algorithm}
\usepackage{multirow}
\usepackage{booktabs}
\usepackage{array}
\usepackage{amsthm}
\usepackage{stfloats}
\usepackage{booktabs}
\usepackage{setspace}
\usepackage{gensymb}
\usepackage{url}
\usepackage{subcaption} 
\usepackage{flushend}
\usepackage{lastpage}
\usepackage{caption}
\allowdisplaybreaks[4]
\DeclareMathOperator{\tr}{tr}

\def\BibTeX{{\rm B\kern-.05em{\sc i\kern-.025em b}\kern-.08em
    T\kern-.1667em\lower.7ex\hbox{E}\kern-.125emX}}
\begin{document}

\title{Multiuser MIMO-AFDM Beamforming for ISAC in Doubly Dispersive Channels
\vspace{-0.3cm}
}

\author{
\IEEEauthorblockN{Rang Liu$^{\star}$, Ming Li$^{\dagger}$, A. Lee
Swindlehurst$^{\ddagger}$, Norman Franchi$^{\star}$, and Robert
Schober$^{\star}$}

\IEEEauthorblockA{
$^{\star}$Friedrich-Alexander-Universität Erlangen-Nürnberg (FAU),
Erlangen 91058, Germany\\
$^{\dagger}$Dalian University of Technology, Dalian, Liaoning 116024,
China\\
$^{\ddagger}$University of California, Irvine, CA 92697, USA\\
Email: \texttt{rang.liu@fau.de, mli@dlut.edu.cn, swindle@uci.edu,}\\
\texttt{\{norman.franchi, robert.schober\}@fau.de}
}\vspace{-0.8cm}
}

\maketitle
\begin{abstract}
Integrated sensing and communication (ISAC) in high-mobility channels requires waveform and beamforming designs that are robust to delay-Doppler dispersion. With this in mind, in this paper we study a monostatic multiuser multiple-input multiple-output (MIMO) affine frequency division multiplexing (AFDM) downlink system. We develop a discrete affine Fourier transform (DAFT)-domain model that preserves Doppler-induced inter-bin coupling and derive a data-aided delay-Doppler detector. The expected matched-bin detector signal-to-noise ratio (SNR) is shown to be proportional to a transmit-covariance beampattern, which leads to a detector-SNR-based sector-illumination constraint. The resulting sensing-constrained weighted sum-rate maximization problem is solved using a combined weighted minimum mean squared error (WMMSE) and majorization-minimization (MM) formulation. Simulations show that the proposed AFDM design outperforms its orthogonal frequency division multiplexing (OFDM) counterpart in terms of the rate-sensing tradeoff, robustness to Doppler, and delay-Doppler sensing quality.
\end{abstract}


\section{Introduction}

Orthogonal frequency division multiplexing (OFDM) has been widely adopted as a baseline waveform for integrated sensing and communication (ISAC), owing to its mature implementation and efficient time-frequency processing \cite{SturmProcIEEE2011,LiuJSAC2022}. However, in high-mobility wideband scenarios, doubly dispersive channels destroy subcarrier orthogonality, leading to Doppler-induced inter-carrier interference and delay-Doppler sensing leakage. Affine frequency division multiplexing (AFDM) has recently emerged as a promising alternative chirp-domain waveform for such channels \cite{Rou2026,BemaniTWC2023,RouSPM2024}. By properly tuning its chirp parameters, AFDM provides a structured representation of delay-Doppler dispersion and improves robustness against rapidly varying sensing channels.

Recent AFDM-ISAC studies have focused on pilot-assisted delay-Doppler sensing, AFDM-parameter design, Cram{\'e}r-Rao bound and ambiguity-function analysis, and transmit waveform optimization \cite{BemaniWCL2024,NiTWC2025,ZhangJSAC2026,QianTIM2026}. These studies have shown that properly designed AFDM pilots and parameters can improve delay-Doppler estimation, extend the unambiguous Doppler range, and provide useful insights into sensing accuracy and ambiguity behavior. However, most existing works focus on point-to-point AFDM-ISAC modeling, pilot-level sensing, receiver-side estimation, or single-vector transmit waveform optimization, rather than multiuser multiple-input multiple-output (MIMO) downlink beamforming. A coherent processing interval (CPI)-level multiuser MIMO-AFDM ISAC framework that jointly addresses multiple data streams, Doppler-induced discrete affine Fourier transform (DAFT)-domain inter-bin coupling, block-level communication rates, data-aided target detection, and sensing-constrained beamforming is not available.

Motivated by this gap, this paper studies a monostatic multiuser MIMO-AFDM ISAC downlink system operating in doubly dispersive channels. We develop a CPI-level DAFT-domain communication and sensing model that retains off-grid Doppler-induced inter-bin coupling. For communication, we adopt a block-level achievable-rate metric for the coupled DAFT-domain symbols; for sensing, we derive a data-aided delay-Doppler matched-filter detector using the known AFDM waveform over the CPI, and show that its expected matched-bin detector signal-to-noise ratio (SNR) crucially depends on the transmit beampattern. This relationship leads to a detector-SNR-based sector-illumination constraint. We then formulate a sensing-constrained weighted sum-rate maximization problem and solve it using a combined weighted minimum mean squared error (WMMSE) and majorization-minimization (MM) formulation with convex quadratically constrained quadratic programming (QCQP) beamformer updates. Simulations demonstrate improved rate-sensing tradeoffs, robustness to Doppler, and delay-Doppler sensing performance compared to OFDM.

\emph{Notation}: The transpose, complex conjugate, and Hermitian transpose operations are denoted by $(\cdot)^T$, $(\cdot)^*$, and $(\cdot)^H$, respectively. The operators $\mathbb{E}\{\cdot\}$, $\operatorname{tr}(\cdot)$, $\operatorname{det}(\cdot)$, $\operatorname{vec}(\cdot)$, $\operatorname{diag}\{\cdot\}$, and $\operatorname{blkdiag}(\cdot)$ represent the expectation, trace, determinant, vectorization, diagonalization, and block-diagonalization functions, respectively. The fields of complex and real numbers are indicated by $\mathbb{C}$ and $\mathbb{R}$, respectively. The circularly symmetric complex Gaussian distribution with mean vector $\boldsymbol{\mu}$ and covariance matrix $\mathbf{C}$ is denoted by $\mathcal{CN}(\boldsymbol{\mu},\mathbf{C})$. For Hermitian matrix $\mathbf{X}$, $\mathbf{X}\succ\mathbf{0}$ indicates that $\mathbf{X}$ is positive definite. The $N\times N$ identity matrix is represented by $\mathbf{I}_N$.

\section{MIMO-AFDM ISAC System Model}
We consider a monostatic AFDM-based ISAC downlink system, where an $N_{\mathrm{t}}$-antenna base station (BS) serves $K$ single-antenna users and probes a point target. The BS is equipped with an $N_{\mathrm{r}}$-element co-located receive uniform linear array (ULA) for monostatic sensing. Each transmitted block of AFDM data carries both information-bearing and dedicated sensing symbols, and the employed precoders must address multiuser communication and sensing requirements.

\subsection{AFDM Transmit Signal Model}
\label{sec:tx_afdm_model}

A given CPI consists of $Q$ AFDM blocks indexed by $q \in \{0,\ldots,Q-1\}$. 
Each AFDM block contains $N$ affine-domain data/sensing symbols that are mapped to $N$ time-domain samples spaced by the baseband sample interval $T_\mathrm{s}$, and is preceded by 
a chirp-periodic prefix (CPP) of length $N_{\mathrm{cpp}}$. The AFDM block duration is therefore $T_{\mathrm{sym}}\triangleq(N+N_{\mathrm{cpp}})T_{\mathrm{s}}$. We assume $N_{\mathrm{cpp}}\ge\ell_{\max}$, where $\ell_{\max} T_s$ denotes the maximum target delay considered in the system.

For DAFT bin $m\in\{0,\ldots,N-1\}$ and AFDM block~$q$, let $\mathbf{d}[m,q]\triangleq[d_1[m,q] \ \cdots \ d_K[m,q]]^{T}\in\mathbb{C}^{K}$ denote the multiuser data symbol vector, and let $\mathbf{c}[m,q]\in\mathbb{C}^{L_{\mathrm{s}}}$ denote the dedicated sensing-symbol vector, where $L_{\mathrm{s}}$ is the number of sensing streams. The data and sensing symbols are zero mean, unit variance, and mutually independent across signal streams, DAFT bins, and AFDM blocks. Hence, $\mathbb{E}\{\mathbf{d}[m,q]\mathbf{d}^H[m,q]\}=\mathbf{I}_K$ and $\mathbb{E}\{\mathbf{c}[m,q]\mathbf{c}^H[m,q]\}=\mathbf{I}_{L_{\mathrm{s}}}$.
The BS applies communication precoder $\mathbf{W}_{\mathrm{c}}[m]\triangleq\big[\mathbf{w}_{\mathrm{c},1}[m],\ldots,\mathbf{w}_{\mathrm{c},K}[m]\big]\in\mathbb{C}^{N_{\mathrm{t}}\times K}$ and sensing precoder $\mathbf{W}_{\mathrm{s}}[m]\in\mathbb{C}^{N_{\mathrm{t}}\times L_{\mathrm{s}}}$ for DAFT bin $m$.
The precoders $\{\mathbf{W}_{\mathrm{c}}[m],\mathbf{W}_{\mathrm{s}}[m]\}_{m=0}^{N-1}$ are allowed to vary across DAFT bins but are kept fixed over one CPI. Thus, the DAFT-domain transmit vector across the $N_{\mathrm{t}}$ antennas is given by
\begin{equation}
\mathbf{s}[m,q]=\mathbf{W}_{\mathrm{c}}[m]\mathbf{d}[m,q]+
\mathbf{W}_{\mathrm{s}}[m]\mathbf{c}[m,q]\in\mathbb{C}^{N_{\mathrm{t}}\times 1}.
\label{eq:tx_precoding}
\end{equation}

Stacking the DAFT-bin transmit vectors row-wise gives
\begin{equation}
\mathbf{S}[q] \triangleq \big[\mathbf{s}[0,q] \; \dots \; \mathbf{s}[N-1,q]\big]^T
\in\mathbb{C}^{N\times N_{\mathrm{t}}} .
\end{equation}
The time-domain AFDM block is generated by applying the inverse DAFT to each transmit-antenna column, i.e.,
\begin{equation}
\mathbf{X}[q] \triangleq \mathbf{A}^H\mathbf{S}[q] \in\mathbb{C}^{N\times N_{\mathrm{t}}},
\label{eq:tx_X_q}
\end{equation}
where $\mathbf{A}\triangleq\boldsymbol{\Lambda}_{c_2}\mathbf{F}\boldsymbol{\Lambda}_{c_1}\in\mathbb{C}^{N\times N}$ is the unitary DAFT matrix, $\mathbf{F}\in\mathbb{C}^{N\times N}$ is the normalized discrete Fourier transform (DFT) matrix, and 
\[
\boldsymbol{\Lambda}_{c_1} \triangleq \operatorname{diag}\{e^{-\jmath 2\pi c_1 n^2}\}_{n=0}^{N-1},
\quad
\boldsymbol{\Lambda}_{c_2} \triangleq \operatorname{diag}\{e^{-\jmath 2\pi c_2 m^2}\}_{m=0}^{N-1}.
\]
Chirp parameters $c_1$ and $c_2$ are chosen according to the delay-Doppler support of the channel. If $c_1=c_2=0$, the model reduces to block-processed OFDM, the CPP becomes the conventional cyclic prefix, and $\mathbf{A}$ the normalized DFT matrix.

\subsection{Doubly Dispersive MIMO Channel Model}
\label{sec:mimo_dd_channel}

Here we derive a generic DAFT-domain representation for a far-field, spatially narrowband, doubly dispersive MIMO channel. For a ULA with half-wavelength spacing, the transmit steering vector for angle $\theta$ is $\mathbf{a}_{\mathrm{t}}(\theta)\triangleq[1 ~e^{-\jmath\pi\sin(\theta)}~ \cdots ~e^{-\jmath(N_{\mathrm{t}}-1)\pi\sin(\theta)}]^T\in\mathbb{C}^{N_{\mathrm{t}}}$, and the receive steering vector $\mathbf{a}_{\mathrm{r}}(\theta)\in\mathbb{C}^{N_{\mathrm{r}}}$ is defined analogously.
Let $\alpha_p\in\mathbb{C}$, $\tau_p\in\mathbb{R}$, and $f_p\in\mathbb{R}$ denote the complex gain, propagation delay, and Doppler shift of channel path $p$, respectively. The corresponding angle of departure (AoD) and angle of arrival (AoA) are denoted by $\theta_{\mathrm{t},p}$ and $\theta_{\mathrm{r},p}$, respectively. The spatial signature of path $p$ is
\begin{equation}
\boldsymbol{\Theta}_p\triangleq
\mathbf{a}_{\mathrm{r}}(\theta_{\mathrm{r},p})\mathbf{a}_{\mathrm{t}}^{H}(\theta_{\mathrm{t},p})
\in\mathbb{C}^{N_{\mathrm{r}}\times N_{\mathrm{t}}}.
\label{eq:chan_Theta_p}
\end{equation}

For monostatic sensing, the transmit and receive angles coincide, i.e., $\theta_{\mathrm{t},p}=\theta_{\mathrm{r},p}=\theta_p$. For a single-antenna user, the receive-side array response reduces to a scalar and is absorbed into the path gain. The path delays are represented by sample-spaced taps of the discrete-time equivalent channel after 
synchronization and sampling, i.e., $\tau_p=\ell_pT_{\mathrm{s}}$ with $\ell_p\in\{0,\ldots,\ell_{\max}\}$. This approximation is reasonable for the considered wideband setting, where the sampling interval provides a fine delay grid \cite{Rou2026}.
In contrast, the Doppler shifts $f_p\in\mathbb{R}$ are retained as continuous physical parameters and are not restricted to the Doppler grid.
The continuous-time, time-varying impulse response between transmit antenna $n_{\mathrm{t}}$ and receive antenna $n_{\mathrm{r}}$ is modeled as
\begin{equation}
h_{n_{\mathrm{r}},n_{\mathrm{t}}}(t,\tau)=\sum_{p=1}^{P_{\mathrm{g}}}
\alpha_p e^{\jmath 2\pi f_p t}[\boldsymbol{\Theta}_p]_{n_{\mathrm{r}},n_{\mathrm{t}}}
\delta(\tau-\tau_p),
\label{eq:chan_tvirf}
\end{equation}
where $P_{\mathrm{g}}$ denotes the number of paths.

After sampling and CPP removal, the sample-spaced delay model preserves the CPP-induced chirp-circular shift structure. Specifically, the integer delay $\ell\leq N_{\mathrm{cpp}}$ is represented by the 
chirp-circular operator $\boldsymbol{\Psi}(\ell)\triangleq\boldsymbol{\Gamma}_{\mathrm{CPP}}(\ell)\boldsymbol{\Pi}^\ell$,
where $\boldsymbol{\Pi}$ is the cyclic-shift matrix and
$\boldsymbol{\Gamma}_{\mathrm{CPP}}(\ell)$ is the unitary diagonal CPP phase-correction matrix for AFDM \cite{BemaniTWC2023}. In AFDM block $q$, the Doppler phase can be decomposed into a slow-time phase across AFDM blocks and a fast-time modulation within each block: $e^{\jmath2\pi f(qT_{\mathrm{sym}}+nT_{\mathrm{s}})}=
b_q(f)e^{\jmath 2\pi f nT_{\mathrm{s}}}$,
where
\begin{subequations}\begin{align}
b_q(f)&\triangleq e^{\jmath 2\pi f qT_{\mathrm{sym}}}, \quad q=0,\ldots,Q-1,  \\
\boldsymbol{\Delta}(f)&\triangleq \operatorname{diag}\big\{e^{\jmath 2\pi f nT_{\mathrm{s}}}
\big\}_{n=0}^{N-1}\in\mathbb{C}^{N\times N}.
\end{align}
\end{subequations}
Thus, the time-domain delay-Doppler operator for one path is
\begin{equation}
\mathbf{T}(\ell,f)\triangleq\boldsymbol{\Delta}(f)\boldsymbol{\Psi}(\ell)\in\mathbb{C}^{N\times N}. 
\end{equation}

Let $\widetilde{\mathbf{Y}}_{\mathrm{g}}[q]\in\mathbb{C}^{N\times N_{\mathrm{r}}}$ denote the received time-domain data block after CPP removal. Combining the temporal delay-Doppler operator and the channel's spatial response yields
\begin{equation}
\widetilde{\mathbf{Y}}_{\mathrm{g}}[q]=\sum_{p=1}^{P_{\mathrm{g}}}
\alpha_p b_q(f_p)\mathbf{T}(\ell_p,f_p)\mathbf{X}[q]\boldsymbol{\Theta}_p^T
+\widetilde{\mathbf{N}}_{\mathrm{g}}[q],
\label{eq:generic_time_mimo_channel}
\end{equation}
where $\mathbf{X}[q]\in\mathbb{C}^{N\times N_{\mathrm{t}}}$ is the block of time-domain transmit data and $\widetilde{\mathbf{N}}_{\mathrm{g}}[q]$ is noise.
Applying the DAFT to each receive-antenna column and using $\mathbf{X}[q]=\mathbf{A}^H\mathbf{S}[q]$, we obtain 
\begin{subequations}\begin{align}
\mathbf{Y}_{\mathrm{g}}[q]&=\mathbf{A}\widetilde{\mathbf{Y}}_{\mathrm{g}}[q] \\
&= \!\sum_{p=1}^{P_{\mathrm{g}}}\!
\alpha_p b_q(f_p)\mathbf{A}\mathbf{T}(\ell_p,f_p)\mathbf{A}^H\mathbf{S}[q]\boldsymbol{\Theta}_p^T
+\mathbf{N}_{\mathrm{g}}[q], 
\end{align}\end{subequations}
where $\mathbf{N}_{\mathrm{g}}[q]\triangleq\mathbf{A}\widetilde{\mathbf{N}}_{\mathrm{g}}[q]$. If $\widetilde{\mathbf{N}}_{\mathrm{g}}[q]$ has independent and identically distributed (i.i.d.) $\mathcal{CN}(0,\sigma^2)$ entries, then so will $\mathbf{N}_{\mathrm{g}}[q]$ due to the unitarity of $\mathbf{A}$. The corresponding DAFT-domain delay-Doppler operator is therefore defined as
\begin{equation}
\mathbf{E}(\ell,f)\triangleq\mathbf{A}\mathbf{T}(\ell,f)\mathbf{A}^H
=\mathbf{A}\boldsymbol{\Delta}(f)\boldsymbol{\Psi}(\ell)\mathbf{A}^H
\in\mathbb{C}^{N\times N}.
\label{eq:chan_E_lf}
\end{equation}
Since $\mathbf{A}$, $\boldsymbol{\Delta}(f)$, and $\boldsymbol{\Psi}(\ell)$ are unitary, $\mathbf{E}(\ell,f)$ is also unitary.
Thus, the corresponding generic DAFT-domain MIMO input-output relationship is given by 
\begin{equation}
\mathbf{Y}_{\mathrm{g}}[q]=\sum_{p=1}^{P_{\mathrm{g}}}
\alpha_p b_q(f_p)\mathbf{E}(\ell_p,f_p)\mathbf{S}[q]\boldsymbol{\Theta}_p^T
+\mathbf{N}_{\mathrm{g}}[q].
\label{eq:generic_daft_mimo_channel}
\end{equation}
Note that \eqref{eq:generic_daft_mimo_channel} retains the full DAFT-domain delay-Doppler operator $\mathbf{E}(\ell_p,f_p)$, thereby capturing the inter-bin coupling induced by off-grid Doppler shifts. The downlink communication and monostatic sensing models used below follow as special cases of this generic model.

\subsection{Multiuser Communication Model and Performance Metric}
\label{sec:downlink_comm}
We now adapt \eqref{eq:generic_daft_mimo_channel} to the multiuser downlink with single-antenna users, where the channel of user $k$ comprises $P_k$ paths parameterized by $\{\alpha_{k,p},\ell_{k,p},f_{k,p},\theta_{k,p}\}_{p=1}^{P_k}$. The received DAFT-domain block at user $k$ during AFDM block $q$ is then given by
\begin{equation}
\mathbf{z}_k[q]=\sum_{p=1}^{P_k}\alpha_{k,p}b_q(f_{k,p})\mathbf{E}(\ell_{k,p},f_{k,p})
\mathbf{S}[q]\mathbf{a}_{\mathrm{t}}^{*}(\theta_{k,p})+\mathbf{v}_k[q],
\label{eq:comm_z_k_q}
\end{equation}
where $\mathbf{v}_k[q]\sim\mathcal{CN}(\mathbf{0},\sigma_{\mathrm{c}}^2\mathbf{I}_N)$ denotes white Gaussian noise at user $k$.

For the subsequent beamforming design, we collect the DAFT-bin symbols and precoders of one AFDM block. Define the stacked user-data vector $\overline{\mathbf{d}}_k[q]\triangleq[d_k[0,q] \ \cdots \ d_k[N\!-\!1,q]]^{T}\in\mathbb{C}^{N}$, and the stacked sensing-symbol vector $\overline{\mathbf{c}}[q]\triangleq[\mathbf{c}^T[0,q] \ \cdots \ \mathbf{c}^T[N\!-\!1,q]]^{T}\in\mathbb{C}^{NL_{\mathrm{s}}}$.
The corresponding block-diagonal communication and sensing precoders are given by
\begin{align}
\overline{\mathbf{W}}_{\mathrm{c},k}&\triangleq\operatorname{blkdiag}(
\mathbf{w}_{\mathrm{c},k}[0],\ldots,\mathbf{w}_{\mathrm{c},k}[N\!-\!1])\!
\in\!\mathbb{C}^{NN_{\mathrm{t}}\!\times\! N},
\label{eq:comm_W_k}\\
\overline{\mathbf{W}}_{\mathrm{s}}&\triangleq\operatorname{blkdiag}(
\mathbf{W}_{\mathrm{s}}[0],\ldots,\mathbf{W}_{\mathrm{s}}[N\!-\!1])\!
\in\!\mathbb{C}^{NN_{\mathrm{t}}\!\times\! NL_{\mathrm{s}}} .
\label{eq:comm_Ws}
\end{align}
Using $\overline{\mathbf{s}}[q]\triangleq\operatorname{vec}\!\left(\mathbf{S}^T[q]\right)$, the stacked transmit vector can be written as
\begin{equation}
\overline{\mathbf{s}}[q]=\sum_{j=1}^{K}\overline{\mathbf{W}}_{\mathrm{c},j}\overline{\mathbf{d}}_j[q]
+\overline{\mathbf{W}}_{\mathrm{s}}\overline{\mathbf{c}}[q]
\in\mathbb{C}^{NN_{\mathrm{t}}\times 1}.
\label{eq:comm_sbar_q}
\end{equation}
By construction, the $m$-th $N_{\mathrm{t}}$-dimensional block of $\overline{\mathbf{s}}[q]$ corresponds to the DAFT-bin transmit vector $\mathbf{s}[m,q]$.
For path~$p$ of user $k$, define the angular projection matrix $\boldsymbol{\Phi}_{k,p}\triangleq\mathbf{I}_N\otimes \mathbf{a}_{\mathrm{t}}^{H}(\theta_{k,p})\in\mathbb{C}^{N\times NN_{\mathrm{t}}}$, which satisfies $\boldsymbol{\Phi}_{k,p}\overline{\mathbf{s}}[q]=\mathbf{S}[q]\mathbf{a}_{\mathrm{t}}^{*}(\theta_{k,p})$.
The effective DAFT-domain channel of user $k$ in block $q$ is then given by
\begin{equation}
\mathbf{H}_k[q]\triangleq\sum_{p=1}^{P_k}\alpha_{k,p}b_q(f_{k,p})
\mathbf{E}(\ell_{k,p},f_{k,p})\boldsymbol{\Phi}_{k,p}\!\in\!\mathbb{C}^{N\!\times\! NN_{\mathrm{t}}} .
\label{eq:comm_H_k_q}
\end{equation}
Thus, \eqref{eq:comm_z_k_q} can be compactly rewritten as
\begin{equation}
\mathbf{z}_k[q]=\sum_{j=1}^{K}\mathbf{H}_k[q]\overline{\mathbf{W}}_{\mathrm{c},j}\overline{\mathbf{d}}_j[q]+\mathbf{H}_k[q]\overline{\mathbf{W}}_{\mathrm{s}}\overline{\mathbf{c}}[q]
+\mathbf{v}_k[q].
\label{eq:zk_stacked}
\end{equation}

Because off-grid Doppler generally makes $\mathbf{E}(\ell_{k,p},f_{k,p})$ dense in the DAFT domain, the effective channel $\mathbf{H}_k[q]$ couples the $N$ DAFT bins. A per-bin signal-to-interference-plus-noise (SINR) problem formulation is therefore not appropriate. We instead adopt a block information-rate metric over the $N$ coupled DAFT-domain symbols.
The interference-plus-noise covariance of user $k$ in block $q$ is given by 
\begin{equation}
\begin{aligned}
\boldsymbol{\Sigma}_k[q]\triangleq &\sum_{j\ne k}
\mathbf{H}_k[q]\overline{\mathbf{W}}_{\mathrm{c},j}\overline{\mathbf{W}}_{\mathrm{c},j}^{H}
\mathbf{H}_k^{H}[q] \\
&+\mathbf{H}_k[q]\overline{\mathbf{W}}_{\mathrm{s}}\overline{\mathbf{W}}_{\mathrm{s}}^{H}
\mathbf{H}_k^{H}[q]+\sigma_{\mathrm{c}}^{2}\mathbf{I}_N .
\end{aligned}
\label{eq:comm_Sigma_k_q}
\end{equation}
Assuming Gaussian signaling and joint decoding over the $N$ coupled DAFT-domain symbols, the achievable block rate of user $k$ is given by 
\begin{equation}
R_k[q]=\log_2\det(\mathbf{I}_N\!+\!
\boldsymbol{\Sigma}_k^{-1}[q]\mathbf{H}_k[q]\overline{\mathbf{W}}_{\mathrm{c},k}
\overline{\mathbf{W}}_{\mathrm{c},k}^{H}\mathbf{H}_k^{H}[q]).
\label{eq:comm_R_k_q}
\end{equation}
The CPI-averaged achievable block rate is then defined as
\begin{equation}
\bar{R}_k\triangleq\frac{1}{Q}\sum_{q=0}^{Q-1}R_k[q],
\label{eq:comm_Rbar_k}
\end{equation}
which serves as the communication metric for the proposed sensing-constrained beamforming design.

\section{AFDM Target Detection and Sensing Metric}
\label{sec:sensing_metric}
This section develops the receiver-side signal processing for monostatic AFDM target detection and derives a detection-SNR metric for transmit beamforming design. Leveraging the known AFDM transmit blocks, the monostatic receiver constructs a data-aided delay-Doppler statistic, whose matched-bin SNR quantifies target detectability. We further show that the expected detector SNR is proportional to the transmit-covariance beampattern, motivating the sensing constraint adopted for beamforming design in Section~IV.

\subsection{Data-Aided Delay-Doppler Matched Filtering}
\label{subsec:target_bin_statistic}

Consider a point target characterized by $\{\alpha,\theta,\ell,f\}$, where $\alpha\in\mathbb{C}$ is the complex reflection coefficient, $\theta$ is the target angle, $\ell$ is the on-grid delay, and $f\in\mathbb{R}$ is the Doppler shift. Since the transmitter and receiver are co-located, the AoD and AoA coincide, and the monostatic spatial response is $\boldsymbol{\Theta}(\theta)\triangleq\mathbf{a}_{\mathrm{r}}(\theta)\mathbf{a}_{\mathrm{t}}^{H}(\theta)\in\mathbb{C}^{N_{\mathrm{r}}\times N_{\mathrm{t}}}$.
From \eqref{eq:generic_daft_mimo_channel}, the received DAFT-domain reflection from a single target is given by
\begin{equation}
\mathbf{Y}[q]=\alpha b_q(f)\mathbf{E}(\ell,f)\mathbf{S}[q]\boldsymbol{\Theta}^T(\theta)
+\mathbf{N}_{\mathrm{s}}[q],
\label{eq:sen_Y_q}
\end{equation}
where $\mathbf{N}_{\mathrm{s}}[q]$ has i.i.d. $\mathcal{CN}(0,\sigma_{\mathrm{s}}^2)$ entries.
Since the transmitted block $\mathbf{S}[q]$ is known at the monostatic receiver, it can be used for data-aided matched filtering over candidate delay-Doppler bins. For a delay-Doppler hypothesis $(\ell',\nu)$, where $\ell'\in\{0,\ldots,\ell_{\max}\}$, and $\nu\in\Omega_\nu$ for sampled Doppler search grid $\Omega_\nu$, the monostatic receiver forms
\begin{equation}
\overline{\mathbf{Z}}[\ell',\nu]=\sum_{q=0}^{Q-1}b_q^*(\nu)
\mathbf{S}^H[q]\mathbf{E}^H(\ell',\nu)\mathbf{Y}[q]
\in\mathbb{C}^{N_{\mathrm{t}}\times N_{\mathrm{r}}} .
\label{eq:slow_time_integration}
\end{equation}
In \eqref{eq:slow_time_integration}, $\mathbf{E}^H(\ell',\nu)$ compensates for the hypothesized DAFT-domain delay-Doppler response, $\mathbf{S}^H[q]$ performs known-waveform matched filtering, and $b_q^*(\nu)$ aligns the slow-time Doppler phase before coherent integration across the $Q$ AFDM blocks. Thus, $\overline{\mathbf{Z}}[\ell',\nu]$ is the virtual MIMO observation associated with the hypothesized delay-Doppler bin.

Define $\mathbf{z}[\ell',\nu]\triangleq\operatorname{vec}\!\left(\overline{\mathbf{Z}}[\ell',\nu]\right)\in\mathbb{C}^{N_{\mathrm{t}}N_{\mathrm{r}}\times 1}$ and the virtual steering vector $\mathbf{a}_{\mathrm{v}}(\theta)\triangleq\operatorname{vec}\!\left(\boldsymbol{\Theta}^T(\theta)\right)=\mathbf{a}_{\mathrm{r}}(\theta)\otimes \mathbf{a}_{\mathrm{t}}^*(\theta)$.
Substituting \eqref{eq:sen_Y_q} into \eqref{eq:slow_time_integration} gives the matched-filter output at a generic delay-Doppler bin as
\begin{equation}
\mathbf{z}[\ell',\nu]=\alpha\left(\mathbf{I}_{N_{\mathrm{r}}}\otimes\boldsymbol{\Xi}[\ell',\nu]
\right)\mathbf{a}_{\mathrm{v}}(\theta)+
\mathbf{z}_{\mathrm{n}}[\ell',\nu],
\label{eq:z_generic_dd_bin}
\end{equation}
where
\begin{equation}
\boldsymbol{\Xi}[\ell',\nu]\triangleq\sum_{q=0}^{Q-1}b_q^*(\nu)b_q(f)
\mathbf{S}^H[q]\mathbf{E}^H(\ell',\nu)\mathbf{E}(\ell,f)\mathbf{S}[q]
\label{eq:Xi_generic_dd_bin}
\end{equation}
is the waveform correlation matrix between the hypothesized and true
delay-Doppler responses, and
\begin{equation}
\mathbf{z}_{\mathrm{n}}[\ell',\nu]=\operatorname{vec}\bigg(
\sum_{q=0}^{Q-1}b_q^*(\nu)\mathbf{S}^H[q]\mathbf{E}^H(\ell',\nu)\mathbf{N}_{\mathrm{s}}[q]
\bigg)
\label{eq:zn_target_bin}
\end{equation}
is the noise term. Conditioned on the transmitted blocks $\{\mathbf{S}[q]\}_{q=0}^{Q-1}$, the covariance of $\mathbf{z}_{\mathrm{n}}[\ell',\nu]$ is obtained as 
\begin{equation}
\mathbf{R}_{\mathrm{n}}\triangleq\mathbb{E}\!\left\{\!
\mathbf{z}_{\mathrm{n}}[\ell',\nu]\mathbf{z}_{\mathrm{n}}^H[\ell',\nu]
\,\big|\, \{\mathbf{S}[q]\}_{q=0}^{Q-1}\!\right\}\!\!=\!
\sigma_{\mathrm{s}}^2\left(\mathbf{I}_{N_{\mathrm{r}}}\!\otimes\!\mathbf{G}\right),\!
\label{eq:Rn_target_bin}
\end{equation}
where we define
\begin{equation}
\mathbf{G}\triangleq\sum_{q=0}^{Q-1}\mathbf{S}^H[q]\mathbf{S}[q].
\label{eq:G_def}
\end{equation} 

The matched-filter output in \eqref{eq:z_generic_dd_bin} provides one virtual MIMO observation for each delay-Doppler hypothesis. Since the front-end search is performed over the delay-Doppler grid only, the target angle is not introduced as an additional search dimension in the sensing map. Instead, it remains in the virtual steering vector $\mathbf{a}_{\mathrm{v}}(\theta)$ and determines the strength of the signal component associated with each delay-Doppler bin.

\vspace{-0.2cm}
\subsection{Detector SNR and Sensing Constraint}
\label{subsec:detection_metric}

The matched-filter outputs in \eqref{eq:z_generic_dd_bin} are converted to a delay-Doppler sensing map through the whitened energy statistic:
\begin{equation} 
T_{\mathrm{DD}}(\ell',\nu)=\mathbf{z}^H[\ell',\nu]\left(\mathbf{R}_{\mathrm{n}}
+\epsilon\mathbf{I}_{N_{\mathrm{t}}N_{\mathrm{r}}}\right)^{-1}\mathbf{z}[\ell',\nu],
\label{eq:TDD_statistic}
\end{equation}
where $\epsilon\ge0$ is a diagonal-loading factor for numerical regularization. A candidate delay-Doppler bin is declared occupied when $T_{\mathrm{DD}}(\ell',\nu)\geq\eta_{\mathrm{FA}}$, where $\eta_{\mathrm{FA}}$ is selected according to the desired false-alarm probability.

We use the matched-bin detector SNR to quantify target detectability. For a matched delay-Doppler hypothesis, i.e., $(\ell',\nu)=(\ell,f)$, we have $\boldsymbol{\Xi}[\ell,f]=\mathbf{G}$, and \eqref{eq:z_generic_dd_bin} reduces to
\begin{equation}
\mathbf{z}[\ell,f]=\alpha(\mathbf{I}_{N_{\mathrm{r}}}\otimes\mathbf{G})\mathbf{a}_{\mathrm{v}}(\theta)+\mathbf{z}_{\mathrm{n}}[\ell,f].
\label{eq:z_matched_single_target}
\end{equation}
For ideal whitening with $\epsilon=0$ and $\mathbf{R}_{\mathrm{n}}\succ0$, the squared norm of the whitened target component, i.e., the noncentrality parameter of the ideal whitened energy detector and the conditional matched-bin detector SNR, is given by
\begin{subequations}\begin{align}
\lambda_{\mathrm{DD}}(\theta)&=|\alpha|^2
\mathbf{a}_{\mathrm{v}}(\theta)^H(\mathbf{I}_{N_{\mathrm{r}}}\!\otimes\!\mathbf{G})^H
\mathbf{R}_{\mathrm{n}}^{-1}(\mathbf{I}_{N_{\mathrm{r}}}\!\!\otimes\!\mathbf{G})
\mathbf{a}_{\mathrm{v}}(\theta)\\
&={N_{\mathrm{r}}|\alpha|^2}{\sigma_{\mathrm{s}}^{-2}}~
\mathbf{a}_{\mathrm{t}}^{T}(\theta)\mathbf{G}\mathbf{a}_{\mathrm{t}}^{*}(\theta),
\label{eq:lambda_DD_def}
\end{align}\end{subequations}
where $\mathbf{R}_{\mathrm{n}}=\sigma_{\mathrm{s}}^2(\mathbf{I}_{N_{\mathrm{r}}}\otimes\mathbf{G})$ and $\mathbf{a}_{\mathrm{v}}(\theta)=\mathbf{a}_{\mathrm{r}}(\theta)\otimes\mathbf{a}_{\mathrm{t}}^*(\theta)$.
For a fixed false-alarm probability, the detection probability is monotone in $\lambda_{\mathrm{DD}}(\theta)$.

The instantaneous SNR in \eqref{eq:lambda_DD_def} depends on the data-aided waveform through $\mathbf{G}$. 
Since the beamformers are fixed over the CPI whereas the transmitted data and sensing symbols are random, we use the ensemble-averaged matched-bin SNR as a deterministic beamforming metric. From the independence and unit variance of the transmitted symbols, we have 
\begin{equation}
\mathbb{E}\{\mathbf{G}\}=Q\mathbf{R}_x^*,
\label{eq:EG_Rx}
\end{equation}
where
\begin{equation}
\mathbf{R}_x\triangleq\sum_{m=0}^{N-1}\bigg(\sum_{k=1}^{K}\mathbf{w}_{\mathrm{c},k}[m]\mathbf{w}_{\mathrm{c},k}^H[m]+\mathbf{W}_{\mathrm{s}}[m]\mathbf{W}_{\mathrm{s}}^H[m]
\bigg)
\label{eq:Rx_from_precoders}
\end{equation}
is the aggregated transmit covariance over the DAFT bins. Since $\mathbf{R}_x$ is Hermitian, taking the expectation of \eqref{eq:lambda_DD_def} yields
\begin{equation}
\mathbb{E}\!\left\{\lambda_{\mathrm{DD}}(\theta)\right\}=
{N_{\mathrm{r}}Q|\alpha|^2}{\sigma_{\mathrm{s}}^{-2}}\mathbf{a}_{\mathrm{t}}^{H}(\theta)
\mathbf{R}_x\mathbf{a}_{\mathrm{t}}(\theta).
\label{eq:Elambda_DD_Rx}
\end{equation}
Thus, the beamformer-dependent part of the expected target-bin detector SNR is the transmit-covariance beampattern $p_\mathrm{sen}(\theta)\triangleq\mathbf{a}_{\mathrm{t}}^{H}(\theta)\mathbf{R}_x\mathbf{a}_{\mathrm{t}}(\theta)$.
As the target direction is only known to lie in the sensing sector $\Theta_{\mathrm{sen}}$, we impose the detector-SNR requirement over a sampled angular grid 
$\mathcal{G}_\theta=\{\theta_1,\ldots,\theta_{N_\theta}\}\subset\Theta_{\mathrm{sen}}$: 
\begin{equation}
p_{\mathrm{sen}}(\theta_i)=\mathbf{a}_{\mathrm{t}}^{H}(\theta_i)\mathbf{R}_x\mathbf{a}_{\mathrm{t}}(\theta_i)\ge\Gamma_{\mathrm{s}},\qquad i=1,\ldots,N_\theta.
\label{eq:sector_sensing_constraint}
\end{equation}
Equivalently, this constraint ensures that the ensemble-averaged target-bin detector SNR is lower bounded in each sampled direction $\theta_i$ by $\gamma_\mathrm{req} = N_{\mathrm{r}}Q|\alpha|^2\Gamma_{\mathrm{s}}\sigma_{\mathrm{s}}^{-2}$.

\section{Sensing-Constrained Beamforming Design}
\label{sec:beamforming}

\subsection{Problem Formulation}

Under the detection-SNR-based sensing constraint derived in
Section~\ref{sec:sensing_metric}, we jointly optimize the communication and sensing beamformers to maximize the weighted sum rate:
\begin{subequations}
\label{eq:P_W_original}
\begin{align}
\max_{\mathcal{W}}\quad&\sum_{k=1}^{K}\mu_k\bar{R}_k(\mathcal{W})\label{eq:P_W_original_obj}\\
\mathrm{s.t.}\quad & \mathbf{a}_{\mathrm{t}}^H(\theta_i)\mathbf{R}_x(\mathcal{W})\mathbf{a}_{\mathrm{t}}(\theta_i)
\ge\Gamma_{\mathrm{s}}, \quad i=1,\ldots,N_\theta, 
\label{eq:P_W_original_sensing}\\
&\tr\big(\mathbf{R}_x(\mathcal{W})\big)\le P_{\max},
\label{eq:P_W_original_power}
\end{align}
\end{subequations}
where $\mathcal{W}\triangleq\{\mathbf{w}_{\mathrm{c},k}[m],\mathbf{w}_{\mathrm{s},\ell_{\mathrm{s}}}[m],\forall k,\ell_{\mathrm{s}},m\}$ collects all beam vectors, and $\mathbf{w}_{\mathrm{s},\ell_{\mathrm{s}}}[m]$ denotes the $\ell_{\mathrm{s}}$-th column of $\mathbf{W}_{\mathrm{s}}[m]$.
The covariance $\mathbf{R}_x(\mathcal{W})$ is given in \eqref{eq:Rx_from_precoders}, $\bar{R}_k(\mathcal{W})$ is the CPI-averaged rate of user $k$, and $\mu_k\ge0$ is its weight. Constraint \eqref{eq:P_W_original_sensing} enforces the required sensing illumination power level $\Gamma_{\mathrm{s}}$, while \eqref{eq:P_W_original_power} imposes the total transmit-power budget $P_{\mathrm{max}}$. Problem \eqref{eq:P_W_original} is non-convex because the user rates are coupled by multiuser interference and the sensing constraints are super-level quadratic constraints.

\subsection{WMMSE-MM Beamformer Optimization}
\label{subsec:wmmse_mm}
We solve \eqref{eq:P_W_original} by combining a WMMSE reformulation of the
weighted sum rate objective function and an MM inner approximation of the non-convex sensing constraints. 

\subsubsection{WMMSE-based Transformation}
We first rewrite the communication objective in a WMMSE form. For user $k$ in AFDM block $q$, define the total receive covariance matrix 
\begin{equation}
\mathbf{C}_k[q]\triangleq\boldsymbol{\Sigma}_k[q]+
\mathbf{H}_k[q]\overline{\mathbf{W}}_{\mathrm{c},k}
\overline{\mathbf{W}}_{\mathrm{c},k}^H\mathbf{H}_k^H[q],
\label{eq:Ck_total_alg}
\end{equation}
which includes the desired signal, multiuser interference, dedicated sensing streams, and noise. For a linear receiver $\mathbf{U}_k[q]\in\mathbb{C}^{N\times N}$, the MSE matrix for estimating the desired DAFT-domain symbol vector of user $k$ is 
\begin{align}
\mathbf{E}_k[q]&=\mathbf{U}_k^H[q]\mathbf{C}_k[q]\mathbf{U}_k[q]
-\mathbf{U}_k^H[q]\mathbf{H}_k[q]\overline{\mathbf{W}}_{\mathrm{c},k}\notag\\
&\qquad-\overline{\mathbf{W}}_{\mathrm{c},k}^H\mathbf{H}_k^H[q]\mathbf{U}_k[q]
+\mathbf{I}_N .
\label{eq:MSE_matrix_alg}
\end{align}
By further introducing a positive definite weighting matrix $\mathbf{M}_k[q]\succ\mathbf{0}$ and using $-\ln\det(\mathbf{E}) = \max_{\mathbf{M}\succ\mathbf{0}}\{\ln\det(\mathbf{M})-\operatorname{tr}(\mathbf{ME})+N\}$ \cite{QShiTSP2011}, the block rate is equivalently written as
\begin{equation}\begin{aligned}
R_k[q]=\max_{\mathbf{U}_k[q],\,\mathbf{M}_k[q]\succ\mathbf{0}}~&
\frac{1}{\ln2}\big[\ln\det(\mathbf{M}_k[q])\\
&\quad~ -\tr\left(\mathbf{M}_k[q]\mathbf{E}_k[q]\right)
+N\big].
\label{eq:WMMSE_identity_alg}
\end{aligned}\end{equation}

Since the constraints involve only $\mathcal{W}$, the auxiliary variables can be updated in closed form at a feasible reference point $\mathcal{W}^\mathrm{ref}$:
\begin{subequations}\label{eq:WMMSE_aux_update_alg}
\begin{align}
\mathbf{U}_k[q]&=
(\mathbf{C}_k^\mathrm{ref}[q])^{-1}\mathbf{H}_k[q]\overline{\mathbf{W}}_{\mathrm{c},k}^\mathrm{ref},
\label{eq:U_update_alg}\\
\mathbf{E}^\mathrm{ref}_k[q]&=\mathbf{I}_N-(\overline{\mathbf{W}}_{\mathrm{c},k}^\mathrm{ref})^H
\mathbf{H}_k^H[q]\mathbf{U}_k[q],\label{eq:E_update_alg}\\
\mathbf{M}_k[q]&=(\mathbf{E}^\mathrm{ref}_k[q])^{-1},
\label{eq:M_update_alg}
\end{align}
\end{subequations}
where $\mathcal{W}^\mathrm{ref}$ contains the feasible beamformers available at the beginning of the update and all quantities with superscript ``ref'' are evaluated using $\mathcal{W}^\mathrm{ref}$.

For fixed auxiliary variables, maximizing the WMMSE surrogate over $\mathcal{W}$ is equivalent to minimizing 
\begin{equation}
\sum_{k=1}^{K}\frac{\mu_k}{Q\ln2}\sum_{q=0}^{Q-1}
\tr\big(\mathbf{M}_k[q]\mathbf{E}_k[q]\big),
\label{eq:weighted_mse_obj}
\end{equation}
where $\mathbf{E}_k[q]$ is obtained with \eqref{eq:MSE_matrix_alg} using the candidate beamformers $\mathcal{W}$ and the fixed receive filter $\mathbf{U}_k[q]$.
Expanding \eqref{eq:weighted_mse_obj} and dropping constants independent of $\mathcal{W}$ gives a convex quadratic objective:
\begin{align}
&f_{\mathrm{WMMSE}}(\mathcal{W};\mathcal{W}^\mathrm{ref})=\sum_{j=1}^{K}
\tr\big(\overline{\mathbf{W}}_{\mathrm{c},j}^H\mathbf{T}_\mathrm{w}\overline{\mathbf{W}}_{\mathrm{c},j}
\big)\notag\\
&\qquad +\tr\big(\overline{\mathbf{W}}_{\mathrm{s}}^H\mathbf{T}_\mathrm{w}\overline{\mathbf{W}}_{\mathrm{s}}
\big)-2\sum_{k=1}^{K}\Re\big\{\tr\big(\mathbf{B}_k^H\overline{\mathbf{W}}_{\mathrm{c},k}
\big)\big\},
\label{eq:wmmse_quad_obj_alg}
\end{align}
where
\begin{subequations}
\label{eq:T_B_def_alg}
\begin{align}
\mathbf{T}_\mathrm{w}&\triangleq\sum_{k=1}^{K}\!\frac{\mu_k}{Q\ln2}\!\sum_{q=0}^{Q-1}\!
\mathbf{H}_k^H[q]\mathbf{U}_k[q]\mathbf{M}_k[q]\mathbf{U}_k^H[q]\mathbf{H}_k[q],
\label{eq:T_def_alg}\\
\mathbf{B}_k&\triangleq\frac{\mu_k}{Q\ln2}\sum_{q=0}^{Q-1}\mathbf{H}_k^H[q]\mathbf{U}_k[q]\mathbf{M}_k[q].
\label{eq:B_def_alg}
\end{align}
\end{subequations}

\subsubsection{MM-based Transformation}
We next minorize the non-convex sensing constraints. Let $\mathbf{a}_i\triangleq\mathbf{a}_{\mathrm{t}}(\theta_i)$ and rewrite the transmit beampattern as
\begin{equation}
g_i(\mathcal{W})\triangleq\mathbf{a}_i^H\mathbf{R}_x(\mathcal{W})\mathbf{a}_i
=\sum_{\mathbf{w}\in\mathcal{W}}|\mathbf{a}_i^H\mathbf{w}|^2.
\label{eq:gi_vector_sum_alg}
\end{equation}
Although $g_i(\mathcal{W})$ is a convex quadratic in the beam vectors, the constraint $g_i(\mathcal{W})\ge\Gamma_{\mathrm{s}}$ is non-convex. For each $\mathbf{w}\in\mathcal{W}$ with corresponding reference vector $\mathbf{w}^\mathrm{ref}$, the first-order lower bound of $|\mathbf{a}_i^H\mathbf{w}|^2$ is given by
\begin{equation}
|\mathbf{a}_i^H\mathbf{w}|^2\ge2\Re\left\{\mathbf{w}^H\mathbf{a}_i\mathbf{a}_i^H
\mathbf{w}^\mathrm{ref}\right\}-|\mathbf{a}_i^H\mathbf{w}^\mathrm{ref}|^2 .
\label{eq:MM_single_term_alg}
\end{equation}
Applying this bound to all beam vectors yields the affine minorant:
\begin{equation}\label{eq:gtilde}
\widetilde{g}_i(\mathcal{W};\mathcal{W}^\mathrm{ref})=\!\!\sum_{\mathbf{w}\in\mathcal{W}}
\!\!\left[2\Re\!\left\{\!\mathbf{w}^H\mathbf{a}_i\mathbf{a}_i^H\mathbf{w}^\mathrm{ref}\!\right\}
-|\mathbf{a}_i^H\mathbf{w}^\mathrm{ref}|^2\right].\!
\end{equation}
This minorant is tight at $\mathcal{W}^\mathrm{ref}$ and satisfies  $g_i(\mathcal{W})\ge\widetilde{g}_i(\mathcal{W};\mathcal{W}^\mathrm{ref})$, thus enforcing $\widetilde{g}_i(\mathcal{W};\mathcal{W}^\mathrm{ref})\ge\Gamma_{\mathrm{s}}$ yields an inner approximation of the original sensing constraint.

\subsubsection{Beamformer Update}
Combining the WMMSE objective and the lower-bounded sensing constraints, the beamformer update is obtained by solving
\begin{subequations}
\label{eq:P_WMMSE_MM_alg}
\begin{align}
\min_{\mathcal{W}}\quad & f_{\mathrm{WMMSE}}(\mathcal{W};\mathcal{W}^\mathrm{ref})\label{eq:P_WMMSE_MM_obj_alg}\\
\mathrm{s.t.}\quad & \widetilde{g}_i(\mathcal{W};\mathcal{W}^\mathrm{ref})\ge\Gamma_{\mathrm{s}},
\quad i=1,\ldots,N_\theta, \label{eq:P_WMMSE_MM_sensing_alg}\\
&\tr\left(\mathbf{R}_x(\mathcal{W})\right)\le P_{\max},
\label{eq:P_WMMSE_MM_power_alg}
\end{align}
\end{subequations}
which is a convex QCQP that can be readily solved by standard optimization tools.

In summary, starting from a feasible solution $\mathcal{W}^\mathrm{ref}$, each iteration updates the auxiliary variables based on  \eqref{eq:WMMSE_aux_update_alg}, constructs the MM minorants with \eqref{eq:gtilde}, solves \eqref{eq:P_WMMSE_MM_alg}, and sets its optimizer as the new reference beamformers for the next iteration. The iterations terminate when the relative improvement of the true weighted sum rate falls below a prescribed tolerance. Since the WMMSE reformulation and MM minorants are tight at the current iterate, the algorithm preserves feasibility and monotonically improves the weighted sum rate. Following standard WMMSE/MM arguments \cite{QShiTSP2011}, it converges to a stationary point of \eqref{eq:P_W_original}.

\begin{figure}
    \centering
    \begin{subfigure}{0.49\linewidth}
        \centering
            \includegraphics[width=\linewidth]{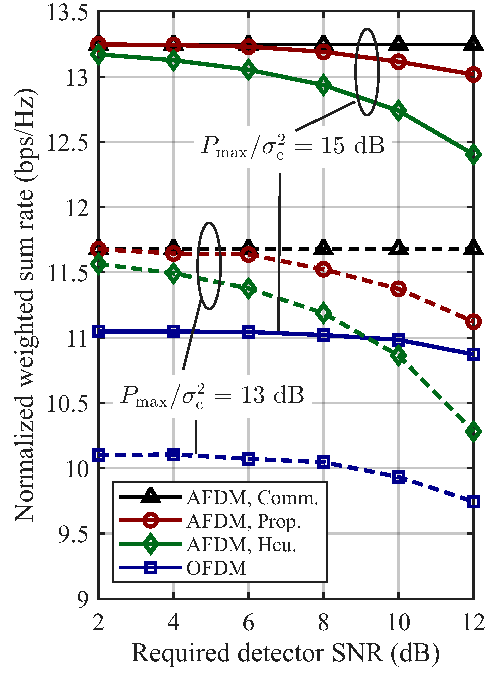}%
            \vspace{-0.2cm}
        \caption{Required detector SNR.}
        \label{fig:rate_sensing_tradeoff}
    \end{subfigure}
    \hfill
    \begin{subfigure}{0.49\linewidth}
        \centering
            \includegraphics[width=\linewidth]{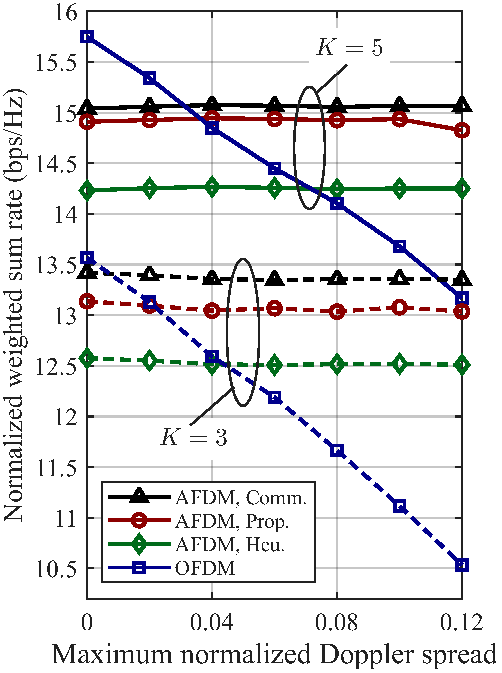}%
            \vspace{-0.2cm}
        \caption{Maximum Doppler spread.}
        \label{fig:rate_velocity}
    \end{subfigure}

    \caption{Weighted sum rate under different settings.}
    \label{fig:rate_tradeoff_combined}
\end{figure}

\section{Simulation Results}

In this section, we evaluate the proposed MIMO-AFDM ISAC design in a high-mobility multiuser scenario. Unless otherwise specified, we set $N_{\mathrm{t}}=N_{\mathrm{r}}=6$, $N=64$, $Q=16$, $\ell_{\max}=8$, $N_{\mathrm{cpp}}=10$, $P_{\max}/\sigma_{\mathrm{c}}^2=15$ dB, and $\mu_k=1,~\forall k$. The $K=3$ single-antenna users are located at $\{-45^\circ,-10^\circ,20^\circ\}$. Each communication channel contains three delay-Doppler paths with gains distributed as $\mathcal{CN}(0,1/3)$; their delays are drawn from $\{0,\ldots,\ell_{\max}\}$, and their normalized Doppler shifts are uniformly distributed over $[-\nu_{\max},\nu_{\max}]$ with $\nu_{\max}=0.1$. The sensing sector is $\Theta_{\mathrm{sen}}=[30^\circ,50^\circ]$, with a target at $\theta_{\mathrm{t}}=40^\circ$. 
The chirp parameters are set as $c_1 = (2\nu_{\mathrm{max}}+1)/(2N)$ and $c_2 = 1/(2N)$.
We compare the proposed joint design, denoted by \textbf{AFDM, Prop.}, with three benchmarks: \textbf{OFDM}, which uses the same optimization framework with $c_1=c_2=0$; \textbf{AFDM, Heu.}, which applies a sequential heuristic algorithm that first designs the sensing beams and then optimizes the communication beams; and \textbf{AFDM, Comm.}, which removes the sensing constraints and serves as a communication-only upper bound.

The rate-sensing tradeoff is shown in Fig.~\ref{fig:rate_sensing_tradeoff}.
As the required detector SNR $\gamma_{\mathrm{req}}$ increases, stricter sensing requirements force more spatial resources to be steered toward the sensing sector, thereby reducing the weighted sum rate. The proposed AFDM scheme remains closest to the communication-only benchmark and consistently outperforms both OFDM and the heuristic AFDM design. The gain over OFDM demonstrates the robustness of AFDM against Doppler-induced dispersion, while the gain over the heuristic design highlights the benefits of jointly optimizing the communication and sensing beamformers.
Fig.~\ref{fig:rate_velocity} evaluates the impact of the maximum normalized Doppler spread $\nu_{\max}$.
For small $\nu_{\max}$, the channel is nearly quasi-static and OFDM remains competitive. As $\nu_{\max}$ increases, however, OFDM suffers from Doppler-induced inter-carrier interference, whereas AFDM achieves a higher rate by better matching the doubly dispersive channel. Consequently, the AFDM gain becomes more pronounced in the high-mobility regime.

\begin{figure}
    \centering
    \includegraphics[width=0.99\linewidth]{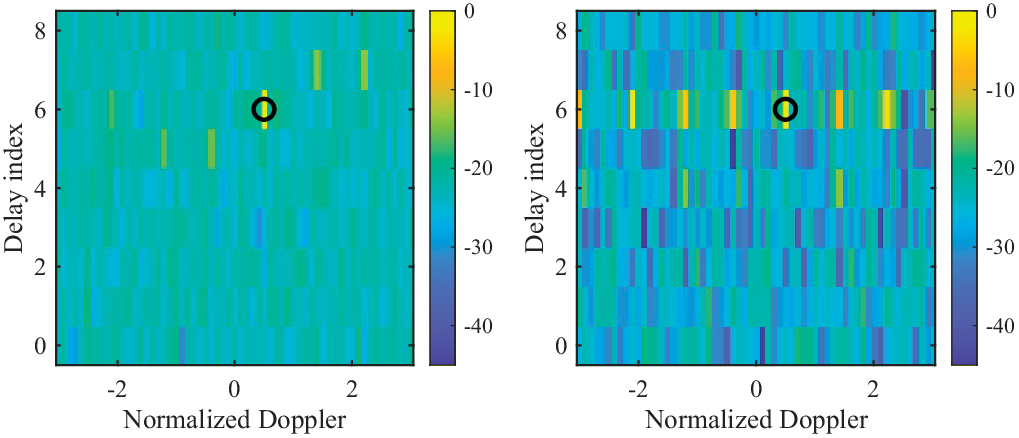}
    \caption{Normalized delay-Doppler detection maps. (Target: black circle. Left: AFDM; right: OFDM).}
    \label{fig:2Dmap}
\end{figure}

Fig.~\ref{fig:2Dmap} compares the normalized delay-Doppler detection maps. Although AFDM and OFDM satisfy the same target-illumination requirement, AFDM produces a more concentrated target peak with a lower sidelobe floor. In contrast, OFDM exhibits stronger Doppler-induced leakage, which raises the background level and makes the target less distinguishable. These results confirm that AFDM improves both the communication-sensing tradeoff and the reliability of data-aided delay-Doppler sensing in doubly dispersive channels.

\section{Conclusion}
We have investigated sensing-constrained beamforming in the monostatic multiuser MIMO-AFDM ISAC downlink with doubly dispersive channels. 
By linking the expected matched-bin detector SNR to the transmit beamforming, we derived a detector-SNR-based sector-illumination constraint and solved the resulting weighted sum-rate maximization via a WMMSE-MM algorithm with convex QCQP beamformer updates. 
Simulations confirmed the enhanced robustness of AFDM over OFDM in high-mobility doubly dispersive channels, motivating future extensions to more complex sensing scenarios.



\begin{thebibliography}{99}

\bibitem{SturmProcIEEE2011}
C. Sturm and W. Wiesbeck, ``Waveform design and signal processing aspects for fusion of wireless communications and radar sensing,'' \emph{Proc. IEEE}, vol. 99, no. 7, pp. 1236--1259, Jul. 2011.

\bibitem{LiuJSAC2022}
F. Liu \textit{et al.}, ``Integrated sensing and communications: Toward dual-functional wireless networks for 6G and beyond,'' \emph{IEEE J. Sel. Areas Commun.}, vol. 40, no. 6, pp. 1728--1767, Jun. 2022.

\bibitem{Rou2026} H. S. Rou, V. Savaux, Z. Sui, G. T. F. Abreu, and Z. Liu, ``AFDM: Evolving OFDM towards 6G+,'' Feb. 2026. [Online]. Available: https://arxiv.org/abs/2602.08163

\bibitem{BemaniTWC2023}
A. Bemani, N. Ksairi, and M. Kountouris, ``Affine frequency division multiplexing for next generation wireless communications,'' \emph{IEEE Trans. Wireless Commun.}, vol. 22, no. 11, pp. 8214--8229, Nov. 2023.

\bibitem{RouSPM2024}
H. S. Rou \textit{et al.}, ``From orthogonal time-frequency space to affine frequency-division multiplexing: A comparative study of next-generation waveforms for integrated sensing and communications in doubly dispersive channels,'' \emph{IEEE Signal Process. Mag.}, vol. 41, no. 5, pp. 71--86, Sep. 2024.

\bibitem{BemaniWCL2024}
A. Bemani, N. Ksairi, and M. Kountouris, ``Integrated sensing and communications with affine frequency division multiplexing,'' \emph{IEEE Wireless Commun. Lett.}, vol. 13, no. 5, pp. 1255--1259, May 2024.

\bibitem{NiTWC2025}
Y. Ni, P. Yuan, Q. Huang, F. Liu, and Z. Wang, ``An integrated sensing and communications system based on affine frequency division multiplexing,'' \emph{IEEE Trans. Wireless Commun.}, vol. 24, no. 5, pp. 3763--3779, May 2025.

\bibitem{ZhangJSAC2026}
F. Zhang \textit{et al.}, ``AFDM-enabled integrated sensing and communication: Theoretical framework and pilot design,'' \emph{IEEE J. Sel. Areas Commun.}, vol. 44, pp. 310--324, 2026.

\bibitem{QianTIM2026}
J. Qian, Y. Liang, Y. Zou, X. Song, and Y. Huang, ``The transmit design for ISAC system based on affine frequency-division multiplexing,'' \emph{IEEE Trans. Instrum. Meas.}, vol. 75, Art. no. 6502711, 2026.

\bibitem{QShiTSP2011}
Q. Shi, M. Razaviyayn, Z. -Q. Luo, and C. He, ``An iteratively weighted MMSE approach to distributed sum-utility maximization for a MIMO interfering broadcast channel,'' \emph{IEEE Trans. Signal Process.}, vol. 59, no. 9, pp. 4331--4340, Sep. 2011.


\end{thebibliography}
\end{document}